\documentclass[runningheads,a4paper]{llncs}

\usepackage{amssymb}
\usepackage{graphicx}
\usepackage{braket}
\usepackage{mathtools, amsmath}
\newtheorem{teorema}{Theroem}

\usepackage{url}
\urldef{\mailsa}\path|enrico.prati@cnr.it|    
\newcommand{\keywords}[1]{\par\addvspace\baselineskip
\noindent\keywordname\enspace\ignorespaces#1}

\begin{document}

\mainmatter  

\title{Quantum Compiling}

\titlerunning{Quantum Compiling}

%
%
\author{Marco Maronese\inst{1,2,3}
\and Lorenzo Moro\inst{1,4,5} 
\and Lorenzo Rocutto\inst{1,2,3}
\and Enrico Prati\inst{1,5}
\thanks{MM, LM and LR equally contributed to this work as first Author. Email: enrico.prati@cnr.it}
}
\authorrunning{M. Maronese et al.}
%
\institute{Quantum Team - Istituto di Fotonica e Nanotecnologie, Consiglio Nazionale delle Ricerche, Piazza Leonardo da Vinci 32, I-20133 Milano, Italy. \\
\and Dipartimento di Informatica - Scienza e Ingegneria - DISI, Alma Mater Studiorum - Università di Bologna, Via Zamboni 33, I-40126 Bologna, Italy.\\
\and Istituto Italiano di Tecnologia, Via Morego 30, I-16163 Genova, Italy.\\
\and Dipartimento di Elettronica, Informazione e Bioingegneria, Politecnico di Milano, Via Colombo 81, I-20133 Milano, Italy. \\
\and Consorzio Interuniversitario delle Telecomunicazioni, Viale G.P. Usberti, 181/A Pal.3, I-43124 Parma, Italy.
}

%
%

\toctitle{Book Title}
\tocauthor{Quantum Compiling}
\maketitle

\begin{abstract}
Quantum compiling fills the gap between the computing layer of high-level quantum algorithms and the layer of physical qubits with their specific properties and constraints. Quantum compiling is a hybrid between the general-purpose compilers of computers, transforming high-level language to assembly language and hardware synthesis by hardware description language, where functions are automatically synthesized into customized hardware.
Here we review the quantum compiling stack of both gate model quantum computers and the adiabatic quantum computers, respectively. The former involves low level qubit control, quantum error correction, synthesis of short quantum circuits, transpiling, while the latter involves the virtualization of qubits by embedding of QUBO and HUBO problems on constrained graphs of physical qubits and both quantum error suppression and correction. Commercial initiatives and quantum compiling products are reviewed, including explicit programming examples.
\end{abstract}

\section{Introduction}

The Chapter is entirely dedicated to the compiling stack of quantum computers. Although the topic has been developed at several layers and by multiple approaches during the history of quantum computation, it has never been reviewed systematically. 

In computer science, the firmware is a class of computer software aimed at providing the low-level control of a specific hardware, in order to enable hardware independence. 
A compiler is a computer program aimed at translating computer code between two languages, called the source and the target, respectively\cite{grune2012modern}. Usually it translates from a high-level programming language to a lower level language (like assembly language) so that the latter can be executed.
For the purpose of this Chapter, we may simplify by saying that the compiler applies an algorithm to generate a firmware. 

High level quantum algorithms require error-free qubits and logic gates, so the main purpose of a quantum compiler is twofold: translating ideal quantum gate operations used in quantum algorithms into machine level operations, and, because of the special nature of quantum computers, to fight against the loss of quantum information during time because of decoherence \cite{hu2002decoherence}.
Because of the complexity of managing quantum systems in practice, the compiling process of the quantum firmware for quantum computer requires a number of stacked operations.
In general we may refer to quantum compiling as classical software algorithms needed to connect the physical operations on a quantum hardware\cite{ladd2010quantum}, which may range from semiconductor\cite{rotta2017quantum} or superconductor chip\cite{huang2020superconducting}, to system based on trapped ions\cite{bruzewicz2019trapped}, or neutral atoms\cite{saffman2019quantum}, with the source code of a high-level quantum algorithm written in terms of error-free quantum logic gates. 
A quantum computer should be seen as the quantum version of a co-processor. Similarly to a graphical processing unit (GPU), it requires a classical chip with classical software to be exploited. The quantum processing unit (QPU) is called by those part of the code involving a quantum algorithm.
The task of the full quantum compiler stack is made complex by its twofold role. Addressing the quantum firmware to enable hardware-independence requires both to map constrained physical operation into high level gates, and to organize groups of physical qubits so to behave collectively as error-free logical qubits, respectively. In this Chapter, we discuss in details how the two aspects are managed by following the layered architecture of quantum computers~\cite{jones2012layered}.
Such layered architecture and the field of quantum compiling have been developed originally by targeting the gate-model quantum computer (like those of IBM, Rigetti, IonQ). Here, we extend the approach to adiabatic quantum computers \cite{albash2018adiabatic}. Instead, one-way qumodes-based \cite{raussendorf2001one} and topological quantum computers \cite{freedman2003topological} are in a too early stage to be included in the analysis. Similarly to the ISO/OSI stack, which conceptualizes the different layers of a network, both gate model and adiabatic quantum computers can be abstracted by stacking distinct layers with different roles connecting the hardware with the quantum algorithm level. 

While the compiling of gate-model quantum computers involves synthesis of quantum gates at both the physical and the logical layers, adiabatic quantum computers require embedding methods to increase the limited connectivity of physical qubits. 
In the following, the layered architecture of quantum computers is introduced for the gate model quantum computer as well as its extension to adiabatic quantum computer. Next, a Section is dedicated to the quantum compiling techniques for gate model quantum computers, another section to those for embedding in adiabatic quantum computers, and finally a section is dedicated to discuss and compare commercial products developed to support the improvement of performances of quantum computers.    

\section{Layered architecture and quantum compiling stack}
In this section, we introduce the concept of the layered architecture of quantum computers. We outline how each layer is separately addressed, and we explain how this architecture, originally conceived for gate-model quantum computers, can be extended to adiabatic quantum computers.

\subsection{The five layers architecture of quantum computers}
To clarify the kind and the encapsulation layer at which the quantum compiling operates, the most straightforward starting point is by introducing the layered architecture of quantum computers.
As highlighted, a quantum processing unit works as a co-processor operated by a classical computer. Therefore, the layered architecture only defines the stack of such a QPU. The main advantage of a modular architecture consists of its hierarchical organization thanks at least nominally to a conceptual separation among different kinds of operations, which are supposed to intervene with some order. Like the TCP/IP OSI/ISO layered architecture~\cite{zimmermann1980osi}, which is adequate for textbooks but is systematically violated in practice, the layered architecture of quantum computers may be seen as a conceptualization of different kinds of operations, that can be more relaxed than the rigid layers would suggest. Therefore, the layers should be considered as a tool to abstract one's architecture rather than some constraint to adapt the design. 

The layered architecture of quantum computers was introduced in 2012 as a conceptual framework for the specific implementation~\cite{jones2012layered} of optically controlled semiconductor quantum dots. Independently of such specific implementation, the architecture has been developed as a general tool suitable for any hardware technology. 

The layered architecture for quantum computers physical design consists of five layers, where each one has a prescribed set of tasks to accomplish. An interface separates all adjacent layers to provide services from the lower layer to the one above it. Ideally, to execute an operation, a layer must issue commands to the layer below and process the results. Some procedures can still be in principle operated between non-adjacent layers. 

The most substantial constraint of the hierarchical layered architecture of Ref.~\cite{jones2012layered} 
is the synchronization of time operation, keeping in mind that the lowest layer will inevitably be asynchronous, being the time evolution of qubits driven by a time-dependent Hamiltonian grounded in a physical hardware. As each layer should output instructions to layers below in a specific sequence, and handling errors is inescapable, a control loop must manage the overall system time evolution. In parallel, syndrome measurements are processed to correct errors. The advantage of the layered architecture for quantum engineers is to focus on individual challenges within an overall design.

\begin{figure}
\centering
\includegraphics[height=6.2cm]{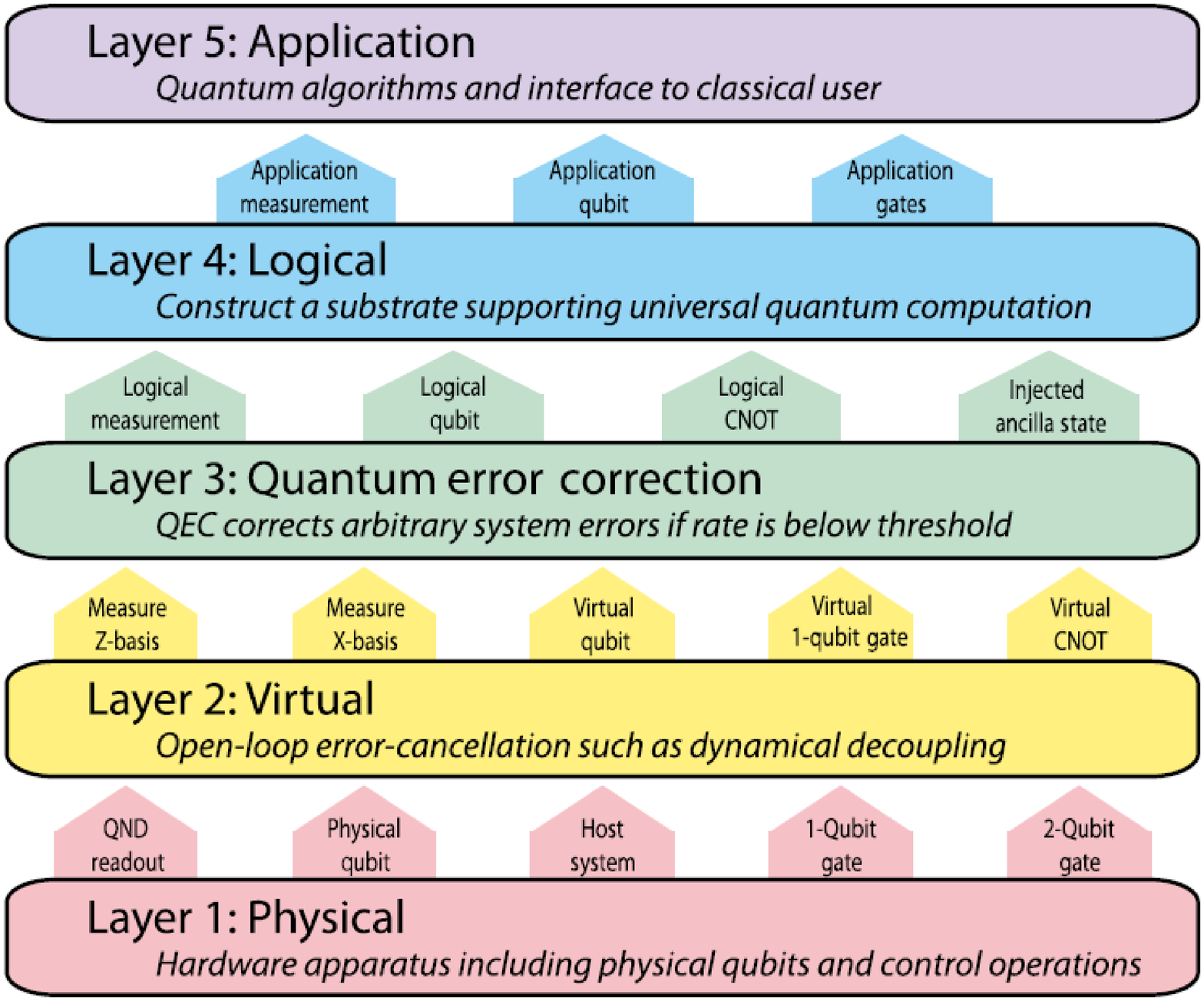}
\caption{The five layers of the architecture of quantum computers. The architecture can be naturally extended to adiabatic quantum computers. The architecture is grounded in Layer 1 Physical. Next, from the bottom to the top, there are: the Layer 2 Virtual, the Layer 3 Quantum Error Correction, the Layer 4 Logical and finally the Layer 5 Application. The latter is not discussed in this Chapter as it has to do with high level algorithms only, while compiling is not involved. Reproduced with permission under the Licence Creative Commons 3.0 from JONES, N. Cody, et al. Layered architecture for quantum computing. Physical Review X, 2012, 2.3: 031007 doi:10.1103/PhysRevX.2.031007.}
\label{fig:example}
\end{figure}

\subsection{Description of the five layers of quantum computers}
\label{layers}
In the following the five layers are described. 

\textbf{Layer 1 - Physical.} There are several different physical implementations on various host systems of a two-level system suitable to be operated as qubits. The most successful are currently superconductive qubits such as flux qubits used by D-Wave quantum annealer~\cite{harris2010experimental} and transmon qubits by IBM~\cite{gambetta2017building},  semiconductor qubits~\cite{rotta2017quantum} \cite{ferraro2020all}, trapped ion qubits\cite{lekitsch2017blueprint}, and neutral atom qubits~\cite{henriet2020quantum} respectively. Here, we do not discuss the approach of one-way quantum computers based on photons~\cite{gu2009quantum} involving qumodes instead of qubits. The physical layer is devoted to control and measure physical qubits. The methods employed at this layer are hardware dependent, intending to hide the physical properties and inaccuracies from higher levels. The physical layer provides unitary control of a qubit by at least two adjustable degrees of freedom, such as rotation around two axes on the Bloch sphere, by three freely adjustable parameters. In some cases, the natural precession of the qubit in the reference frame of the laboratory can be exploited to limit the control to one degree of freedom only.
For instance, short pulses of either electric, magnetic, or electromagnetic fields (such as microwave pulses for superconductor qubits and lasers for trapped ions qubits\cite{moro2020optical}, respectively) can represent the control of the qubit. As dephasing naturally occurs, such control pulses characterize the error source model and provide some adjustments. The physical layer involves 1-qubit gates, 2-qubit gates, a readout mechanism.

\textbf{Layer 2 - Virtual.} The virtual layer collects the quantum dynamics of qubits and shapes them into virtual qubits and quantum gates. In computer science, a virtual object behaves according to a predetermined set of rules, without
a specified object structure. The typical example is the ready-to-use three-spin qubit~\cite{ferraro2014effective}\cite{ferraro2015effective} forming a robust all-electrically controlled two-level system that acts as the effective qubit, involved in creating logical qubits in the gate model quantum computer by the layers above. In some cases, the virtual qubit is built by a single physical qubit. One of the aims of the virtual layer is to eliminate systematic errors. Compensation sequences at Layer 2 can correct correlated errors due to imperfections in the control operations of the gates in Layer 1~\cite{tomita2010multi}, such as electronic noise, fluctuations in laser intensity, or the strength of the coupling of a quantum-dot spin to spins or a microwave or a laser. The compensation sequence may work if the correlated errors happen on time scales longer than operations of the chosen architecture. Hence, a compensation sequence is effective, and the virtual gate has a lower net error than each of the constituent gates in the sequence. In the original Ref. \cite{jones2012layered} introducing the 5-layers, the Authors point out that many compensation sequences are quite general, so error reduction works without knowledge of the type or magnitude of the error. On the contrary, if one tunes the time-dependent Hamiltonian of the control operations~\cite{khodjasteh2009dynamically}, the method falls under the name of dynamically corrected gates. One good reason to use artificial intelligence is connected to the need to characterize the accuracy of operations in the Virtual layer.

\textbf{Layer 3 - Quantum error correction.} The quantum error correction (QEC) layer at its maximum degree of effectiveness is supposed to support fault-tolerant quantum computing. QEC is needed because of the insufficient ability to correct errors of Layer 2. Quantum error correction is unfeasible on NISQ hardware for the simple reason that tens of physical qubits need to be allocated to ensure the information of one qubit survives long enough. Typical quantum error correction methods are concatenated Steane code and surface code\cite{devitt2013quantum}. While in Layer 2, the fundamental principle is to make correlated errors to cancel each other, here isolated general errors are removed. Ideally, Layer 3 would complete the hardware-aware section of the stack and present to Layer 4 fault-tolerant qubits and logic gates only. In practice, in the NISQ era, the qubits presented to Layer 4 are not sufficient for fault tolerance, but they can be used for small simulations. Notice that QEC methods such as the mentioned Steane code correct errors in real-time, while topological codes, such as the surface code, track the faulty qubits and recover the answer after being monitored with the platform.

\textbf{Layer 4 - Logical.} The logical layer is the first hardware-independent layer from the bottom to the top. Such independence is supported by the quantum error correction layer, which guarantees fault tolerance by providing error-free gates. Usually, the reasoning is based on gate-model quantum circuits. The need of a logical layer arises from the limited number of gates offered by Layer 3, such as Clifford gates~\cite{bravyi2005universal}. An easy way to see the reason for such a limit is by thinking that a finite number of syndrome qubits imposes a lower limit to the resolution to distinguish a rotation gate error. Layer 4 compiles all the possible unitary gates given the limited set of error-free quantum gates provided by Layer 3. For a gate-model quantum computer implementing surface codes, the logical layer will provide logical Pauli frames (i.e the stored Pauli gates to be applied to correct the corresponding error at the end of the computation), distillation of ancilla states, the full Clifford group (if this is provided partially from Layer 3), and the approximation of arbitrary quantum gates. 

\textbf{Layer 5 - Application.} The application layer is hardware independent and relies on either the logical (for the gate model) or the virtual (for the adiabatic quantum computer) qubits as arranged by the management of the layers below. This is a high-level programming layer as the classical computer code delivers the algorithm as a sequence of high-level operations, consisting of a quantum circuit in the gate-model quantum computer or the embedding with the annealing schedule on the adiabatic quantum computer. The algorithms consist of the mathematical flow involving the qubits to achieve either a deterministic or probabilistic result. The quantum firmware is applied to the high-level algorithm elements to be translated into physical operations of the quantum hardware. Still, it is not directly involved in the quantum compiling itself. Layer 5 can be used to determine the error correction strategy and the number of hardware resources to be collected from the low-levels.




\subsection{Transposing the five layers architecture to adiabatic quantum computers}

Optimization problems require to find the global minimum of a certain cost function, that can be seen as an energy function. A well known classical approach to solve optimization problems is Simulated Annealing (SA). In SA, we make use of thermal fluctuations to let the system overcome energy barriers standing between the actual state and the ground state for the energy function. At higher temperatures, exploration is fast. Temperature is then slowly lowered so that the system is forced into a minimum, which hopefully corresponds to the lowest energy achievable.
    
It is interesting to wonder if such a paradigm can be extended to a quantum mechanical system. Consider a mechanical system with an associated energy function that we want to minimize using SA. One can introduce artificial degrees of freedom of quantum nature, which in turn introduce quantum fluctuations. In principle, quantum fluctuations can be considered as the tendency of the quantum system to explore the phase space of the classical configurations. We can initially set a high amplitude for such quantum fluctuations, so to make the system eager to explore. Then the strength has to gradually decrease to finally vanish as the system hopefully moves to the ground state. In this way, we are using quantum fluctuations to mimic the effects that temperature has in SA.

An algorithm that controls a quantum system by implementing a schedule where quantum fluctuations strength is gradually reduced is called a Quantum Annealing (QA) algorithm. The physical idea underlying such a procedure is to keep the system close to the instantaneous ground state of the quantum system, analogously to the quasi-equilibrium state to be kept during the time evolution of SA. In QA, quantum tunneling between different states replaces thermal hopping in SA.

QA is a generic algorithm applicable, in principle, to any combinatorial optimization problem and is used as a method to reach an approximate solution within a given finite amount of time. Although SA is usually considered a useful and effective method for solving such problems, evidence exists that QA can outperform SA in certain cases. In reference \cite{farhi2002quantum}, the authors compare the required running time for SA and QA in some optimization problems. The problem consists of searching for the optimal configuration among a finite set of configurations each represented by $n$ bits. The authors show that there exist problems for which QA running time is polynomial in $n$ and SA running time grows more than polynomially in $n$. 

Despite this capability, QA has yet to find useful practical applications. Indeed, the major drawback of QA is that a full practical implementation should rely on a quantum computer since time-dependent Schr{\"o}dinger equations with a very large scale have to be solved. Unfortunately, quantum computers are still at an early stage of development, and only small-size problems can be effectively solved. Nonetheless, QA theory suggests that future quantum devices could tackle problems considered difficult for SA methods.

In recent years, adiabatic quantum computers (AQCs) have undergone a fast development. Such devices are a physical realization of the QA concept and are currently investigated as an alternative paradigm of quantum computation.

Even if the layered architecture of quantum computers has been developed for the gate model architecture, adapting to the specific features of adiabatic quantum computers is a natural generalization which provides insights of the two methods. In the remaining of this subsection, the adaptation of the layered architecture to adiabatic quantum computers is outlined.

\textbf{Layer 1 - Physical. } 
As the aim of an adiabatic quantum computer is to maintain a many-body quantum system in its ground state while transforming its Hamiltonian from a generally non-interacting to a connected form, the main point of the control of the physical layer consists of preserving the gap between the ground state and excited states. The main methods falls into the category of energy gap protection (EGP) \cite{jordan2006error} and dynamical decoupling (DD) \cite{lidar2008towards}, respectively. 

\textbf{Layer 2 - Virtual. } 
In a planar chip of superconductive qubits the number of connections from each qubits is necessarily limited, currently of the order o $2^{3-4}$. In order to increase the number of connections, the strategy consists of virtualization of qubits by strongly connecting pairs or groups of qubits so they behave as a single qubit. This method is called embedding and can be considered the heart of quantum compiling for an adiabatic quantum computer. 
\cite{rocutto2020quantum}\cite{rocutto2021quantum}.

\textbf{Layer 3 - Quantum error suppression and correction.} 
Adiabatic quantum computers have been considered incompatible with a naive implementation of stabilizer codes \cite{young2013error}. Layer 3 consists mainly of providing error suppressed virtual qubits. Quantum error correction is  possible by considering non-equilibrium dynamics in encoded AQC by cooling local degrees of freedom i.e. qubits \cite{sarovar2013error}.

\textbf{Layer 4 - Logical.} 
Independently from the work operated by the Layer 3, consisting of either error suppression or error correction or both, the logical layer can be exploited to cast more general problems then QUBo problems so to solve Higher-order Unconstrained Binary Optimization (HUBO)  \cite{boros2002pseudo}. 
Hence, the logical qubits are replaced by HUBO embeddings, based on the QUBO embedding addressed at the bottom layers~\cite{boros2002pseudo}. 

\textbf{Layer 5 - Application.} Adiabatic quantum computers are programmed at high level by algorithms of three kind: minimization of functionals, graph partitioning and sampling statistical distributions \cite{rocutto2020quantum}.

\section{Quantum compiling of gate-model quantum computers}
The section addresses the quantum compilation problem for gate-model quantum computers as a crucial step to translate high-level quantum algorithms in terms of elementary operations implementable on real-world quantum hardware.

\subsection{Gate-model quantum computers}
To understand why quantum compilers are fundamental to run quantum algorithms on quantum hardware, we first need to recall how gate-model quantum computers process the information~\cite{rieffel:polak}.

Gate-model quantum computers work, in some aspects, similarly to their classical counterpart. Although classical computers process information via logical and arithmetical operations and gate-model quantum computers by exploiting quantum phenomena such as superposition and entanglement, they both relies on applying transformations on few bits at once to perform computations. Unitary transformations mathematically describe quantum computation. Therefore, unitary matrices acting on the state of two-level physical systems called qubits represent quantum gates. 

To achieve general-purpose quantum computation, we need to build devices that can implement any quantum computation, i.e. any unitary transformation acting on the qubits. However, physical constraints limit quantum computers to apply few set of gates, and quantum errors make every instruction count. Therefore, there is the need for specialized software, i.e. quantum compilers, to provide robust control of the computation and map the quantum computation into ordered sequences of gates implementable on real quantum hardware.

In gate-model quantum computers, quantum compiling synthesizes quantum logic gates at two different layers of the stack. At Layer 2 it consists of mapping arbitrary quantum gates into the constrained unitary operations of the physical qubits of a specific hardware technology. At Layer 4 it consists of mapping the set of quantum gates resulting from the quantum error correction layer, which is usually limited, to arbitrary quantum gate to be made available to Layer 5.    

\subsection{The standard circuit model}
The standard circuit model is one of the first well-known theoretical results in the quantum computation framework to achieve gate model quantum computers~\cite{PhysRevA.52.3457}. According to the model, it is always possible to achieve quantum computation as an ordered sequence of transformations acting on single and two-qubit subsystems, i.e. as a circuit of quantum gates. Although the resulting circuit requires a finite number of gates that manipulate the information locally and entangle the qubits in pairs, it is completely equivalent to the n-qubit computation. 
Here, we outline the steps of the demonstration, without the proof: 
\begin{description}
    \item[First step.] It consist of showing that any unitary transformation $U$ on a d-dimensional Hilbert space can be decomposed as a sequence of \textsc{CNOT} gates and multi-controlled two-level unitary matrices \textsc{C$^n$-U}.
    \item[Second step.] Next the multi-controlled two-level unitary matrices \textsc{C$^n$-U} are decomposed as sequence of \textsc{CNOT} gates and controlled single-qubit unitary matrices \textsc{C-U}.
    \item[Third step.] Finally the controlled single-qubit unitary matrices \textsc{C-U} are written as sequence of single-qubit unitary matrices and \textsc{CNOT} gates.
\end{description}
The standard circuit model seems to simplify building an n-qubit gate model quantum computer considerably. Instead of fabricating a device that controls all the n-qubits at once, it is possible to achieve the computation by manipulating them individually or in pairs by employing \textsc{CNOT} gates. 

However, such a result has a little practical advantage. On the one hand, it requires a device that can implement every possible single-qubit unitary transformation. On the other hand, real quantum hardware cannot implement all single-qubit gates due to quantum noise and fabrication constraints in their architecture. More importantly, its main drawback is the length of the quantum circuit resulting from such a procedure, which scales exponentially with the number of qubits. Such a massive number of gates would lead to an unmanageable  noisy computation since real-world quantum computers can not implement quantum gates exactly due to noise and inevitable non-idealities resulting from the manufacturing.
Additionally, each gate requires a finite time to be executed. Therefore, any quantum algorithm with an exponential number of logical operations to be performed would hardly offer any computational advantage over the classical counterpart. Such problems are highly relevant nowadays, where gate-model quantum computers can rely on a few qubits and little error correction techniques are applicable.

The standard circuit model leaves unsatisfied due to its limited practical value. It would be more convenient to build general-purpose quantum computers that can implement few quantum gates rather than all of them. However, a finite set of quantum gates cannot generate any unitary transformations perfectly (unless using infinite length circuits), but to approximate the desired computation within an arbitrary accuracy at most. It is worth noticing that such constraint would not represent a substantial restriction to quantum computations since unavoidable noises would limit the possibility of distinguishing arbitrarily close unitaries anyway. 

How to find a strategy to determine the approximating sequence and understand which sets of gates could be exploited represent the main question addressed by the quantum compiling problem.

\subsection{The quantum compiling problem}
Informally, the quantum compiling problem consists of finding an optimal strategy to map quantum algorithms as circuits of quantum gates chosen by a given set. It is a fundamental problem in quantum computation theory, affecting different layers of abstraction, such as the logical and physical layers, depending on the set of gates taken into account and specific hardware constraints to be satisfy (see Section~\ref{layers}). 

At a high-level of abstraction, quantum compilers are usually exploited as quantum transpilers. The tasks typically consist of expressing the circuits into a different set of gates with a "similar level of abstraction" or optimizing a quantum circuit. In such a context, quantum transpilers can increase the performance by reducing the number of ancillae qubits or preferring some particularly optimized gates to decrease the run-time and the overall noise. Some hardware constraints can be taken into account, as the limited connectivity of the qubits. However, the resulting circuits are typically further compiled in later stages, as they are expressed by a high-level set of gates such as the Clifford+T library~\cite{Gottesman_1998}.

At a low-level of abstraction, the quantum compilers take an arbitrary quantum algorithm (equivalently a quantum circuit expressed using high-level quantum gates) and approximate it, within a given tolerance, as a sequence of low-level transformations that can be implemented directly on quantum hardware such as the \textsc{R/XX} library on trapped-ions quantum computers~\cite{linke2017experimental} or \textsc{U1}, \textsc{U2} and \textsc{U3} on \textsc{IBM} \textsc{QX} architectures~\cite{zulehner2019compiling,larose2019overview}.

Regardless of the layer of interest, three key features characterized quantum compilers~\cite{diffusivegates} measuring their performance:
\begin{description}
    \item[Circuit depth.] It represent total number of quantum gates in the circuit.
    \item[Pre-compilation time.] It corresponds to the time taken by the compiler to be ready for use. It is usually performed once before exploiting the compiler.
    \item[Execution time.] It is the time that the compiler takes to return the sequence after the pre-compilation phase.
\end{description}
Such quantities must scale optimally as a function of the accuracy requested, i.e. they do not grow exponentially. However, they would hardly scale optimally simultaneously~\cite{diffusivegates}, and we must choose a trade-off between speed and accuracy. 
For instance, the naive strategy of trying every possible sequence of gates would minimize the circuit depth and requires no pre-compilation time. Still, the execution time would explode very rapidly, making it unfeasible as a quantum compilation strategy.
How to find a strategy to determine the approximating sequence, understand which sets of gates could be exploited and balance those features, remained unclear until the Solovay-Kitaev theorem in the late 1990s.
\subsection{The Solovay-Kitaev theorem}
The Solovay-Kitaev theorem~\cite{dawson2005solovay} is a crucial result in quantum computation and a breakthrough in the quantum compilation problem. Robert M. Solovay first announced the results in 1995, but they were formalized and published independently a few years later by Alexei Y. Kitaev in 1997 on a review paper, including an algorithm to quickly approximate quantum gates~\cite{kitaev1997quantum}. 

The theorem roughly states that if we consider any quantum algorithm i.e. an unitary transformation $U \in SU(d)$, it is possible to find very quickly an approximating sequence of gates as long as they belong to a suitable set $\mathcal{B}$. Such set needs to satisfy some requirements to be exploited:
\begin{enumerate}
    \item All the gates in $\mathcal{B}$ needs to be unitary matrix with determinant $1$;
    \item $\forall A_j \in \mathcal{B}$ the inverse operation $A^{\dagger}_{j} \in \mathcal{B}$;
    \item $\mathcal{B}$ must be an universal set, i.e. it is possible to approximate any unitary operation as a finite sequence of gates from the set.
\end{enumerate}
It is worth highlighting that the second request is not strictly necessary approximate unitary transformations, but it is a condition required to proof the theorem only~\cite{zhiyenbayev:diffusive}.
However, if the requirements are met, the theorem holds for every desired accuracy $\varepsilon$ and the length of the resulting sequence scales efficiently. 
\begin{teorema}[Solovay-Kitaev]
    Let be $\varepsilon>0$ a desired fixed accuracy, then $\forall\, U \in SU(d)$  exists a finite circuit $C$ of gates driven by a set $\mathcal{B}$ such that the distance $d\bigl(U,C\bigr)<\varepsilon$. The sequence $S$ of gates has a length $O\bigl(\log^{c}{(1/\varepsilon)}\bigr)$, where $c$ is a constant value.
\end{teorema}
The circuit depth returned by the theorem and the execution-time of its implementations scale polylogarithmically, even if distinct formulations and proofs achieve different values of the constants~\cite{barenco1995elementary,chuang:efficient}. 
For instance, the Dawson-Nielsen formulation~\cite{dawson2005solovay} provides sequences of length $\mathcal{O}(\log^{3.97}(1/\varepsilon))$ in a time of $\mathcal{O}(\log^{2.71}(1/\varepsilon))$, while in~\cite{kitaev2002classical} those quantities scales as $\mathcal{O}(\log^{3+\gamma}(1/\varepsilon))$ and $\mathcal{O}(\log^{3+\gamma}(1/\varepsilon))$ respectively, where $\gamma$ is a positive constant which can be set at will. Additional boosts in performance can be gained by introducing some constrains such as restricting $\mathcal{B}$ a to diffusive set of gates~\cite{zhiyenbayev:diffusive} the space of unitary matrices efficiently or by employing ancilla qubits~\cite{kitaev2002classical}.

Geometrical proof~\cite{chuang:efficient} shows that despite of the strategy considered, no algorithm can return sequence using less than $\mathcal{O}(\log(1/\varepsilon))$ gates.
Moreover, one critical issue that the theorem does not address is finding a universal set of gates, but fortunately, it can be proved that almost all sets of gates have such propriety~\cite{lloyd:almostanyuniversal,ekert:universality}.

\subsection{Beyond the Solovay-Kitaev theorem}
The Solovay-Kitaev theorem provides an elegant and efficient classical algorithm for compiling an arbitrary unitary transformation into a circuit of quantum gates, balancing the sequence length, the pre-compilation time and the execution time. However, algorithms based on the theorem do not represent the only potential strategies to approach the quantum compiling problem. 

An optimal quantum circuit for a general two-qubit gate requires at most 3 CNOT gates and 15 elementary one-qubit gates. In case of a a purely real unitary two-qubit gate transformation, the construction requires at most 2 CNOTs and 12 one-qubit gates \cite{PhysRevA.69.032315}. Such method, known as KAK decomposition, is used for instance by the IBM QX architectures to address two-qubits random $SU(4)$ transformations.\cite{zulehner2019compiling}  
For instance, the Quantum Fast Circuit Optimizer (Qfactor) optimizes the distance between a sequence of unitary gates, and a target unitary matrix, using an analytic method based on the SVD operation.
In the following, modern approaches are outlined.

\subsubsection{Machine learning approach to quantum compiling}
Machine learning and artificial intelligence approaches have been recently proposed as an alternative strategy~\cite{cao2019quantum}. 
Compiling of arbitrary unitaries into a sequence of gates native to a quantum processor has been for instance obtained by an A* inspired algorithm. Such algorithm is conceived for Noisy-Intermediate-Scale Quantum devices era, as it aims to minimize the number of CNOT used while accounting for connectivity \cite{davis2020towards}. 
Although machine learning approaches can return great quality results and optimized circuits, they usually have high execution-times due to the limited pre-compilation steps that can be employed. In contrast, deep learning approaches exploiting artificial neural networks, which mimic human brains and can be trained like any other machine learning algorithm, seems to be of great promise. By exploiting a neural network, which is trained on a pre-compilation stage beforehand, it would be possible to considerably speed-up the execution-time. 
Such networks can be trained in a supervised fashion~\cite{Swaddle_2017} by a generated training data set containing optimal circuits only. The neural network, once trained, can decompose a unitary matrix into a product of quantum gates. The approach is limited by the quality of the data-set and the size of the neural network, which scales sub-optimally in the number of qubits.

Deep reinforcement learning\cite{porotti:CTAP,porotti2019reinforcement,paparelle2020digitally} can represent an alternative approach \cite{An_2019}. The basic idea is to train a reinforcement learning agent to learn a suitable policy to approximate unitary transformations. The agent is not told how to learn such a policy, but it has to learn through interactions with an environment. It is a time-consuming task, but it has to be performed once in the pre-compilation stage.\cite{moro2021} Although such a strategy could return a short circuit in minimal time, there is no guarantee that the agent will always find it.
Hybrid approaches, where a planning algorithm such as A* is boosted by deep neural networks, could achieve even better performance~\cite{PhysRevLett.125.170501}. However, the execution time is raised by the planning algorithm, which could scale sub-optimally for high accuracy.

\subsubsection{Hardware dependent quantum compilers}
Circuital quantum computers are based on several architectures and topology connections between qubits \cite{larose2019overview,Linke3305,debnath2016demonstration}, both due to the different physical two-level systems exploited and the difficulty connecting qubits. Therefore, low-level quantum compilers are asked to consider the particular topology of a quantum computer architecture to map quantum circuits efficacy into instruction that can be run on quantum hardware.

\begin{figure}
\centering
\includegraphics[width=\textwidth]{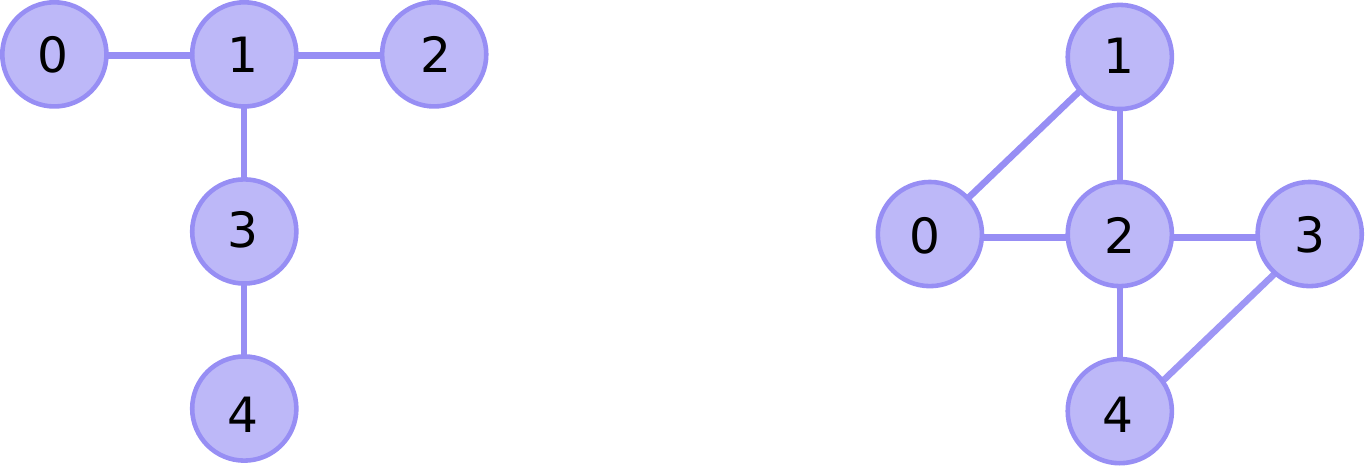}
\caption{Different topologies for the IBM QX 5-qubit quantum computers. Ourense, Valencia and Vigo share the same connectivity (left side), which is more limited than Melbourne (right side). Low-level quantum circuits have to meet hardware constraints: while it is possible to apply directly a \textsc{CNOT} gate between gates $1$ and $3$ on Valencia, it is impossible on Melbourne.}
\label{fig:arch}
\end{figure}

For many physical realizations of quantum computers, the interaction distance between gate qubits is performed between adjacent qubits only, i.e., between nearest neighbors. A potential strategy to overcome the qubits limited connectivity is to ensure that the circuit satisfies a linear-nearest-neighbor \textsc{LNN} structure. In such a scheme, every two-qubit transformation must act on physically adjacent qubits. Although it is possible to meet the nearest neighbor constraints very quickly in a linear time by adding in front of each gates \textsc{SWAP} gates in a "cascade fashion", such naive strategy implies a considerable increase in the circuit depth and therefore in the execution time. Many low-level quantum transpiler strategies have been proposed to decrease the number of additional \textsc{SWAP} gates exploiting a global~\cite{Wille2014ExactRO} or a local~\cite{saeedi2011synthesis,hirata2009efficient,wille2016look} reordering scheme. 

\begin{figure}
\centering
\includegraphics[width=\textwidth]{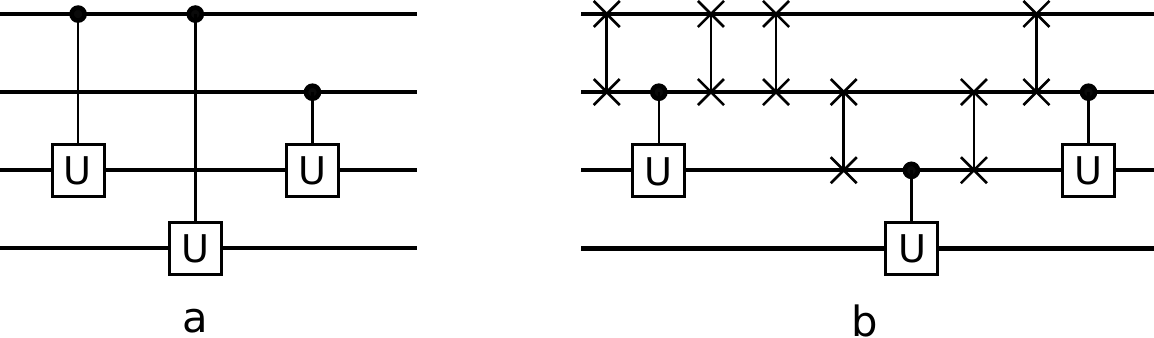}
\caption{Example of circuit transpilation to meet LNN constraint. A circuit that doesn't meet the \textsc{LNN} constraint can be transpiled into a new circuit where every two-qubit gate acts on physically adjacent qubits by adding \textsc{SWAP} gates.}
\label{fig:circ}
\end{figure}

Additional approaches exploit different strategies and machine learning techniques~\cite{wille2016look} to map the circuits into physical ones for specific architectures such as IBM QX architecture or particular unitary sets~\cite{zulehner2019compiling}. 

IBM's approach, which is implemented in its own SDK QISKit, is based on Bravyi's algorithm. It has limited performance mainly because it relies on random searches to meet the physical constraints. In contrast, look-ahead schemes \cite{zulehner2018efficient,wille2016look}, which consider gates applied in the near future and exploit additional information on the circuit, explore a larger part of the search space, leading to increased performance.

\section{Quantum firmware for adiabatic quantum computers}
The layered architecture of quantum computers can be used to conceptualize the different kind of quantum firmware and quantum control for making robust adiabatic quantum computers. In this section, the methods developed to address the five layers are discussed.

\subsection{Layer 1 -- Physical: the annealing process}
    \label{sec:the annealing process}
    
    The starting point consists of considering an optimization problem that can be represented as the ground-state search of a spin--glass model of the general form
    \begin{equation}
    \label{eq:isinggenericHam}
        H_P\equiv -\sum_{i=1}^Nh_i\sigma_i^z-\sum_{i<j}J_{ij}\sigma_i^z\sigma_j^z\;,
    \end{equation}
    where the $\sigma_i^z$ are the Pauli matrices that act along the $z$-direction. Many combinatorial optimization problems can be written in this form, by mapping binary variables to spin variables.
    
    To realize Quantum Annealing (QA), a fictitious kinetic energy is typically introduced by the time-dependent transverse field
    \begin{equation}
        H_{T}\equiv-\sum_{i=1}^N\sigma_i^x\;.
    \end{equation}
    Each term $\sigma_i^x$ allows spin flips, quantum fluctuations, or quantum tunnelling between the states that possess eigenvalues +1 and -1 with respect to $\sigma_i^z$. Such effects allow a quantum search of the phase space. The total Hamiltonian takes the expression
    \begin{equation}
    \label{eq:isingmodelannealing}
    \begin{split}
        H(t)&=-F(t)\left(\sum_{i<j}J_{ij}\sigma^z_i\sigma^z_j+\sum_ih_i\sigma^z_i\right)-G(t)\sum_i\sigma^x_i\\ &\equiv F(t)H_P+G(t)H_T\;,
    \end{split}
    \end{equation}
    where $t$ is the physical time. The parameters $J_{ij}$ and $h_i$, that identify the problem to be solved, can be manually set by the user. Usually $J_{ij}$ are called \textit{couplings} or \textit{weights}, while $h_i$ are referred to as \textit{biases}. The explicit time-dependence of functions $F$ and $G$ can also be controlled, up to a certain freedom. The shape of both $F$ and $G$ together is referred to as \textit{annealing schedule}.
    
    It is important to note that Eq. \ref{eq:isingmodelannealing} introduces a Hamiltonian $H_P$ with only linear and quadratic terms with respect to the spin matrices. The reason for this restriction comes from the fact that modern quantum annealers can only implement Hamiltonians with interaction terms that are at most quadratic.
    
    Each eigenstate of $H(\tau)$ is a set of $N$ binary values $S = \{s_1,s_2...s_N\}$, where $N$ is the total number of spin degrees of freedom of the system. Each spin variable $s_i$ can assume two different states, $+1$ and $-1$. There are $2^N$ different eigenstates of $H_P$, one for each possible combination of the $N$ spin variables.
    
\begin{figure}
        \centering
        \includegraphics[width=0.7\columnwidth]{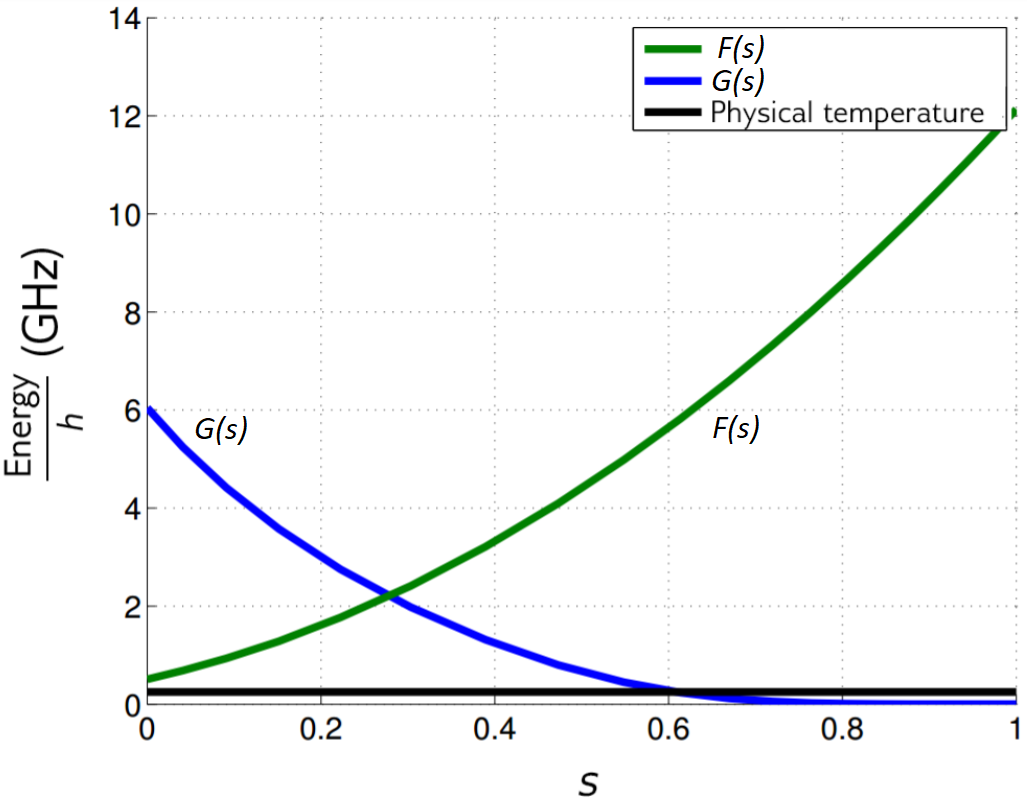}
        \caption{Annealing functions $F(s)$, $G(s)$. Data shown are representative of D-Wave 2X Systems. Image from Ref. \cite{dwave2019technical}.}
        \label{fig:FGannealingdwavemanual}
    \end{figure}

    Suppose we are interested in finding which of the eigenstates corresponds to the minimum energy of $H_P$. By choosing a correct annealing schedule for $F(t)$ and $G(t)$, we can increase the probability to make the system to converge to the global minimum. At $t=0$, we want $G(0)\gg F(0)$ since the system should be allowed to be in any superposition of the classical states. Indeed, in the limit $G(t)\to\infty$ the system should find itself in an eigenstate of Hamiltonian $H_T$, which means each configuration $S$ has an equal probability. As $t$ grows, we raise $F(t)$ and lower $G(t)$. The process gradually forces the system into states that are a mixture of low-energy configurations with respect to Hamiltonian $H_P$. For $t\to\infty$, $G(t)\ll F(t)$, so that the system finds itself in a state superposition which is dominated by the state corresponding to the spin configuration that minimizes $H_P$. If the annealing schedule is sufficiently slow, the adiabatic theorem grants that the system will remain close to the lowest energy eigenstate of the instantaneous Hamiltonian $H(t)$. The Reader may refer to Refs.  \cite{morita2008mathematical,morita2007convergence} for the details of  the mathematical foundations of quantum annealing and the adiabatic theorem. Figure \ref{fig:FGannealingdwavemanual} shows actual time schedules for $F(t)$ and $G(t)$ used in a D-Wave quantum annealing device.
    
\subsubsection{Physical realization of AQCs}
    
    The Josephson junction is the building block of quantum annealing devices. It is built from two pieces of superconducting metal, separated by a \emph{weak link}, which means a thin layer of normal metal or some type of insulator. 
    
    When a superconductor is cooled under the critical temperature $T=T_C$, the electrons in the superconductor form pairs, called Cooper pairs. Each Cooper pair is to be considered a boson, and thus all electrons can condense in the ground energy level. It follows that all the electrons can be described by a collective wave-function having a single quantum phase:
    \begin{equation}
        \Psi(\vec{r},t) = \lvert\Psi(\vec{r},t)\rvert e^{i\varphi(\vec{r},t)}\;.
    \end{equation}
    
    Since such macroscopic wave function must be single-valued in going once around a superconducting loop, flux quantization arises. It means that the flux contained in a closed superconducting loop takes values that are multiples of the flux quantum $\Phi_0=h/2e\approx2.07\times 10^{-15}$ Wb. Here $h$ is the Planck's constant and $e$ is the electronic charge.
    
    The D-Wave QPUs are built with a network of small quantum circuits composed by one or more Josephson junctions embedded in a superconducting closed loop. Such circuits are called radio-frequency superconducting quantum–interference device (rf-SQUID). Thanks to the quantization of the currents inside the Josephson junction, it has been proven that rf-SQUIDs can be modeled as a two states quantum system. The two basis states for the quantum system can be chosen so they have opposite magnetic moment, resulting in a spin up and a spin down state. Interestingly, SQUIDs have been one of the first macroscopic object showing quantum superposition effects \cite{friedman2000quantum}. See Ref. \cite{zimmerman1966macroscopic} for one of the first articles about the SQUID technology.
    
    The Josephson junction, if properly inserted into superconducting loops, allows the construction of the couplers that provide the quadratic terms in $H_P$ (Eq. \ref{eq:isingmodelannealing}). For more on the principles of couplers implementation see Ref. \cite{harris2009compound}. On the other hand, biases in Eq. \ref{eq:isingmodelannealing} can be realized by applying an external magnetic flux on the qubit.
    
\subsubsection{Sources of error in an AQC}

    Although the control parameters $h_i$ and $J_{ij}$ in Eq. \ref{eq:isingmodelannealing} are specified by the user as double-precision floats, some loss of fidelity occurs in implementing these values in the D-Wave QPU. Specifically, instead of finding low--energy states to the optimization problem defined by $H_P$, the QPU solves a slightly altered problem that can be modeled as: 
    
    \begin{equation}
        H_P^\delta=\sum_{i}^N(h_i+\delta h_i)\sigma^z_i+\sum_{i<j}^N(J_{ij}+\delta J_{ij})\sigma^z_i\sigma^z_j\;,
    \end{equation}
    where $\delta h_i$ and $\delta J_{ij}$ characterize the errors in the parameters $h_i$ and $J_{ij}$, respectively.

    Error $\delta h_i$ depends mainly on $h_i$, on all incident couplings $J_{ij}$ and on neighbour biases $h_j$. In the same way, $\delta J_{ij}$ depends mainly on spin and coupling in the local neighborhood of $J_{ij}$. Forecasting the entity of the errors can be very difficult, since a modification on the value of a single bias or coupling propagates and influences also neighboring control parameters. D-Wave Systems declares that the probability distribution of $\delta h_i$ and $\delta J_{ij}$ is approximately Gaussian, with mean $\mu^h_i$, $\mu^J_{ij}$ and standard deviation $\sigma^h_i$, $\sigma^J_{ij}$. Such values depend on the annealing fraction $s$. The Gaussian distribution is interpreted as the sum of two errors, a systematic contribution $\mu$ and a random component with standard deviation $\sigma$. Such distance from the user--specified parameters is called Integrated Control Error (ICE).

    On top of the errors on the value of the parameters, the annealing process is also affected by a coupling with the system surrounding the qubits, that invalidate the hypothesis of adiabaticity. The main physical source for such external coupling is the thermal bath at some temperature, which causes excitations in the qubits, resulting in undesired spin flips.

    Ever since AQC has gained popularity, researchers have been interested in devising computational techniques to suppress or correct errors in the annealing process \cite{jordan2006error}, \cite{lidar2008towards}, \cite{quiroz2012high}, \cite{young2013error}, \cite{pearson2019analog}, \cite{ayanzadeh2020post}. The term \textit{error suppression} regards techniques that aim to average out the effect of casual and systematic errors, while we talk about \textit{error correction} algorithms when considering procedures that, after the annealing process, spot and correct erroneous behaviours of the system. Error suppression techniques belong to the lowest layers of the quantum computers architecture, while error correcting codes belong to the third layer.

\subsubsection{Cloning the problem}

    Each time the QPU is queried with the request to solve a problem, the control parameters $J_{ij}$ and $h_i$ are re-initialized. Thus, the random part of the ICE for each qubit changes, while the systematic part remains fixed. To increase the resistance against random errors a good approach simply consists in running the annealing cycle multiple times, obtaining a greater number of configurations. Unfortunately, the systematic error depends on physical differences between the qubits, and cannot be compensated by running more cycles.
    
    In general, an algorithm written to run on an AQC does not make use of all the available qubits on the device. If the number of qubits used is sufficiently low, the user is able to clone the problem in different locations of the QPU. If the problem can be implemented in $N$ different areas of the QPU, while avoiding that qubits belonging to different replicas are neighbours, the AQC will be able to generate $N$ samples within a single annealing process. Since each sample now requires $1/N$ of the time previously required, such approach linearly improves the running times. In addition, since the problem is implemented in different locations of the QPU, the systematic component of ICEs can be partially averaged out.
    
\subsubsection{Local cooling and dynamical decoupling}
    
    During the quantum annealing process, thermal effect manifest themselves as entropy injected into the system. Over time, such entropy increases, leading the process to failure within a short time. Techniques have been devised to control the entropy of the system during the quantum annealing process. As an example, Ref. \cite{viola1999dynamical}, introduces the Dynamical Decoupling technique. It consists in using tailored time--dependent perturbations as a tool to improve system performances, aiming to decouple a generic open quantum system from any environmental interaction. On this same line, Ref. \cite{sarovar2013error} propose to couple each qubit to a very low-temperature bath that serves as the entropy sink for the system. 
    
    Despite being theoretically solid, this techniques are nowadays still waiting to be fully applied at the hardware level, and thus their effectiveness has yet to be experimentally evaluated.

\subsection{Layer 2 -- Virtual: AQC topology}

\subsubsection{Embedding as the compiling method for problems on adiabatic quantum computers}

    Current technologies allow the realization of AQCs composed by thousands of qubits.  Nonetheless, the number alone is not sufficiently informative, as the graph of their connections is of importance as well. An optimization problem composed of many binary variables will likely require the implementation of many quadratic terms linking such variables. Such terms appear in a summation in Equation \ref{eq:isingmodelannealing}, but one should be wary that the summation contains only those term for which a physical connection between the two qubits exist inside the device. Such physical connections, realized inside the device thanks to Josephson junctions, are called couplers. The current intensity for each coupler can be set manually by the user and appears in Eq. \ref{eq:isingmodelannealing} as the term $J$.
    
    Topology is considered one of the main bottlenecks to the use of modern AQCs (see, e.g. \cite{dumoulin2014challenges}). The reason for this can be understood considering that for instance in the  AQC produced by D-Wave Systems in the year 2020, Advantage\textsuperscript{TM}, each qubit is connected by a coupler to at most $15$ other qubits. Considering the device has more than $5000$ qubits, it follows that only $~ 0.3\% $ of the possible connections are realized. Such limitation makes it difficult to use the device straightforward to solve real-world problems, where variables influence each other in complex ways. To circumvent such limitation, embedding techniques have been developed to enhance connectivity by creating virtual qubits composed by multiple physical qubits.

\subsubsection{Embedding techniques}
\label{sec: embedding techniques}
\begin{figure}
        \begin{center}
        \includegraphics[width=0.5\columnwidth]{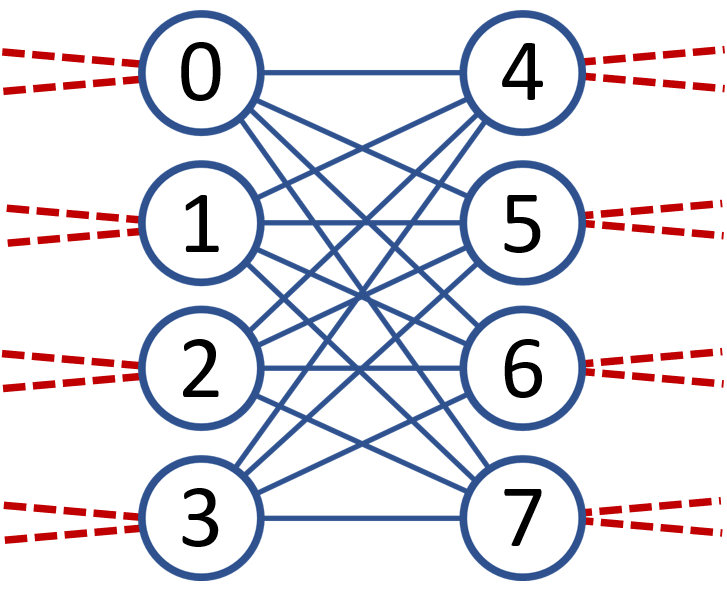}
        \caption{Graph showing existing couplers between qubits inside a cell of a D-Wave 2000Q\textsuperscript{TM} System processor. Red dashed lines represent existing couplers that links the qubits in the cell to qubits in other cells.}
        \label{fig:Chimera_orig}
        \end{center}
    \end{figure}

    Consider the following problem Hamiltonian 
        \begin{equation}
            H_\Delta = J_{1,2}\sigma^z_1\sigma^z_2+J_{1,3}\sigma^z_1\sigma^z_3+J_{2,3}\sigma^z_2\sigma^z_3\;,
        \end{equation}
    Despite involving only three qubits, this problem cannot be submitted to the D-Wave 2000Q\textsuperscript{TM} System processor, since there are no fully connected three-qubits graphs in its topology (see Fig. \ref{fig:Chimera_orig}). Nonetheless we can identify two distinct qubits in a single \emph{virtual qubit}. This means defining a new problem Hamiltonian by adding an additional degree of freedom
    \begin{equation}
    \label{eq:HembDELTA}
            H^\text{emb}_\Delta =H'_\Delta + J_{1_a,1_b}\sigma^z_{1_a}\sigma^z_{1_b}\;,
    \end{equation}
    where $H'_\Delta$ is obtained from Hamiltonian $H_\Delta$ by redefining $\sigma^z_1$ as:
    \begin{equation}
        \sigma^z_1 = \frac{\sigma^z_{1_a}}{2}+\frac{\sigma^z_{1_b}}{2}\;.
    \end{equation}
    If $J_{1_a,1_b} \ll - \max\{\lvert J_{1,2}\rvert,\lvert J_{2,3}\rvert,\lvert J_{1,3}\rvert\}$, it can be concluded that any quantum state where the magnetic moment of qubit $1_a$ and $1_b$ are antiparallel corresponds to a high energy eigenvalue. Thus, such a state is strongly suppressed and does not make a perceivable contribution to the samples statistics. The coupling $J_{chain}$ that links two qubits belonging to the same virtual qubit is usually set to the greatest (negative) value possible, and we will refer to its value as \emph{chain coupling}.
    
    This simple observation allows thinking of qubits $1_a$ and $1_b$ as a single (virtual) unit, since they will almost always agree on their final value, in analogy of how for instance a double spin qubit is used for creating a singlet-triplet (virtual) qubit \cite{fogarty2018integrated}. If we use two qubits instead of one to represent a single unit, a greater number of connections are accessible. Indeed, the problem Hamiltonian $H_\Delta$ cannot be implemented, while $H^\text{emb}_\Delta$ (from Eq. \ref{eq:HembDELTA}) poses no problem. This fact is represented in Fig. \ref{fig:Chimera_orig}. Hamiltonian $H^\text{emb}_\Delta$ can be implemented by making the following identifications:
    \begin{equation}
    \begin{split}
        &\text{Unit } 1_a \rightarrow \text{qubit } 0\\
        &\text{Unit } 1_b \rightarrow \text{qubit } 4\\
        &\text{Unit } 2 \;\rightarrow \text{qubit } 1\\
        &\text{Unit } 3 \;\rightarrow \text{qubit } 5
    \end{split}
    \end{equation}
    
    If such virtual qubits are used, the problem graph is said to have been \emph{embedded} to respect the device restrictions. Inside the D-Wave 2000Q\textsuperscript{TM} System, each qubit is reached by at most 6 couplers. This makes embeddings often necessary, otherwise, most problems could not be submitted to the QA processor. Unfortunately, virtual qubits can behave differently than single qubits during the anneal. The next Section introduces such a concept.
    
\subsubsection{Behaviour of multi-qubit chains}
    \label{sec:Behaviour of multi-qubit chains}
    
    It has been theoretically predicted and experimentally demonstrated that quantum annealing in physical devices produces samples (approximately) as if they were extracted from a Boltzmann distribution at temperature $T_\text{eff}$ (\cite{amin2015searching}, \cite{johnson2011quantum}, \cite{boixo2016computational}, \cite{benedetti2016estimation}). The Boltzmann distribution is well known in statistical mechanics, and in our case it assumes the following aspect:
    
    \begin{equation}
    \label{eq:boltzhypothesischapter4}
        P(S) = \frac{e^{-\frac{E_P(S)}{T_\text{eff}}}}{\sum_{S'\in\Omega}e^{-\frac{E_P(S')}{T_\text{eff}}}}\;,
    \end{equation}
    
    where $S=\{s_i\}$ is a spin configuration belonging to the set of classical states $\Omega$ and $E_p(S)$ is the eigenvalue that the quantum state corresponding to $S$ possesses with respect to the problem Hamiltonian $H_P$.
    
    Such distribution originates from the fact that annealing stops working properly around a certain point $s = s^*$ of the annealing schedule, preventing spins to relax from then on (see Ref. \cite{boixo2016computational}). Such phenomenon is called \textit{freezing}, since spin flips are unlikely to happen after that phase, as if we had lowered the temperature in a simulated annealing algorithm. The resulting effective temperature $T_\text{eff}$ of the distribution directly depends on $s^*$. Indeed, if the annealing could proceed without ever freezing, the final distribution would correspond to Eq. \ref{eq:boltzhypothesischapter4} at an effective temperature $T_\text{eff}\approx0$, thus yielding the correct solution with probability $\approx 1$. In any other case, we have that an early freezing point corresponds to a higher value for $T_\text{eff}$.
    
    Due to such freezing effect, it is fairly easy to find expressions for $H_P$ such that the final distribution obtained through quantum annealing differs sensibly from the expected Boltzmann distribution. Ref. \cite{korenkevych2016benchmarking} shows that it is sufficient to consider a problem where the graph is composed of clusters, i.e. subgraphs with strong ferromagnetic couplings. This situation is customary when the problem is embedded in the physical hardware making use of virtual qubits. If such clusters have different values for their intra-cluster ferromagnetic couplings, they freeze-out at different points during annealing. In general, clusters with great intra-cluster couplings freeze out earlier, thus equilibrating under a Boltzmann distribution at a higher $T_\text{eff}$ than clusters with weak couplings (See also Ref. \cite{amin2015searching}). The result of annealing such a system is a distorted distribution that deviates from the classical Boltzmann distribution.
    
    It follows that one should realize embeddings where each virtual qubit gathers the same number of physical qubits. If that is not the case, virtual qubits composed by longer chains of physical qubits will freeze-out earlier, since they contain an higher number of strong couplings $J_\text{chain}$. If it is not possible to use chains of the same length, it can be useful to synchronize the freeze-out region of each virtual qubit. This is done by delaying by different times the anneal schedule of each virtual qubit, depending on the number of physical qubits it corresponds to. If more physical qubits are coupled together, the anneal schedule should be delayed by more time.

\subsection{Layer 3 -- Quantum error correction} 
\label{sec:Layer 3 -- Error correction in quantum annealing}

\subsubsection{Quantum Annealing Correction (QAC)}
    Ref. \cite{pudenz2014error} introduces an error correcting code to reduce the effect of non-adiabatic transitions on the computation. Such technique is called QAC (quantum annealing correction), and consists in the introduction of an energy penalty in the cost function, along with a particular embedding strategy.
    
    In the first step we encode $H_P$, substituting each $\sigma_i^z$ term with the following encoded counterpart: $\overline{\sigma_i^z}=\sum_{l=1}^n \sigma_{i_l}^z$, so that each logical spin variable is now represented by $n$ distinct spin variables. We also substitute each $\sigma_i^z\sigma_j^z$ with $\overline{\sigma_i^z}\overline{\sigma_j^z}=\sum_{l=1}^n\sigma_{i_l}^z\sigma_{j_l}^z$. We obtain the following encoded Hamiltonian:
    
    \begin{equation}
        \bar{H}_P= \sum_{i<j}^{\bar{N}}J_{ij}\overline{\sigma_i^z}\overline{\sigma_j^z} + \sum_i^{\bar{N}}h_i\overline{\sigma_i^z}\;,
    \end{equation}
    where $\bar{N}$ is the number of encoded qubits. We consider valid only those states for which $\overline{\sigma_i^z}$ has eigenvalues $+n$ or $-n$, which means all qubits representing $\overline{\sigma_i^z}$ are found in the same state.
    
    We are increasing the problem dimension, but in turn we obtain an increased protection against erroneous bit-flips, in two ways. First, the overall problem energy scale is increased by a factor of $n$. The energy gap between the configurations is increased, thus reducing the effect of thermal fluctuations. Second, if the states of the qubits representing the same virtual spin variable
    differ at the end of the annealing process, we can recover the probable correct state by majority vote. This code has minimum Hamming distance $n$, that is, a non-code state with more than $n/2$ bit-flip errors will be incorrectly decoded.
    
    To generate additional protection, the second step of the procedure introduces the following penalty term:
    \begin{equation}
        H_\text{penalty} = - \sum_{i=1}^{\bar{N}}\left( \sigma_{i_1}^z + ... + \sigma_{i_n}^z \right)\sigma_{i_P}^z\;.
    \end{equation}
 
    Such addiction energetically penalizes all bit-flip errors except the full--encoded qubit flip. The role of $H_\text{penalty}$ can be described as to force each problem qubit into agreement with the penalty qubit $\sigma_{i_P}^z$. Indeed, if a qubit does not agree with the penalty qubit, the total sum is increased by 1, raising the energy of the penalty function.

\subsection{Layer 4 -- Logical: Implementing QUBO and HUBO problems on D-Wave}

\subsubsection{QUBO and Ising problems}

    QUBO problems is a class of combinatorial optimization problems that can be solved by a quantum annealing device. The acronym stands for Quadratic Unconstrained Binary Optimization problems. It means that every QUBO problem corresponds to the minimization of a cost function, which is a quadratic expression of binary variables. The keyword \textit{unconstrained} means that there are no constraints that the variables must satisfy a priori (but constraints can be implemented by the user by adding appropriate terms to the cost function).
    
    Every QUBO problem can be written in the form
    
    \begin{equation}
    \label{eq:generic QUBO}
        f(\{s_i\}) = \sum_{i,j}J_{ij}s_is_j+\sum_ih_is_i\;,
    \end{equation}
    with $\{s_i\}$ binary variables and $f$ is the cost function to minimize. In QUBO problems usually $s_i\in \{0,1\}$. The cost function is said to be expressed in its Ising form when the binary variables assume values $-1$ and $1$.
    
    We remind that AQC natively solve problems in the Ising form. Nonetheless, problems are often passed to the AQC in their QUBO form, since the control system of the QPU only needs to perform a simple mapping $s^\prime_i \rightarrow s_i*2 -1$ in Eq. \ref{eq:generic QUBO}, so that $s_i\in\{-1,1\}$. The mapping rescales and shifts the cost function, so such transformation does not influence the nature of the problem nor its solutions. We can conclude that using binary variables in $\{0,1\}$ or $\{-1,1\}$ is just a matter of conventions.
    
\subsubsection{HUBO problems}
    
    As already anticipated in Sec. \ref{sec:the annealing process}, modern AQCs can implement cost function that are at most quadratic. This is a bad news for anyone needing to solve a HUBO problem, which stands for Higher-order Unconstrained Binary Optimization problem. Nonetheless, we just saw in Sec. \ref{sec: embedding techniques} how embedding techniques can be used to implement problems that do not respect the topology of the quantum device. In the same way, an appropriate embedding allows the implementation of HUBO problems on an AQC.
    
    A useful algorithm has been devised in Ref. \cite{boros2002pseudo}. It expresses the following idea: whenever a HUBO problem contains a cubic term, e.g. $-5 s_1 s_2 s_3$, we can introduce a new binary variable $s_{1,2} = s_1s_2$. In this way, the order of the problem is reduced by introducing a new variable. If we repeat the process until no more cubic term is present, we have obtained a QUBO problem. Each new variable must be equal to the product of two already existing variables. This property is enforced by adding a proper \textit{penalty term} to the QUBO cost function. Now, the two following equations hold:
    \begin{equation}
    \label{eq:penalty for HUBO}
    \begin{split}
        &s_{1,2} = s_1 s_2\;\; \text{iff}\;\; C(x,y,z) = 3s_{12}+s_1s_2-2xz-2yz=0\\
        &s_{1,2} \neq s_1 s_2\;\; \text{iff}\;\; C(x,y,z) = 3s_{12}+s_1s_2-2xz-2yz>0
    \end{split}
    \end{equation}
    The function $C(x,y,z)$ has multiple minima, but they correspond to each and every possible combination of values for $x$, $y$, $z$ that respect the relation $z=xy$. It is then sufficient to multiply $C(x,y,z)$ by a parameter $M\gg0$ and then add it to the cost function of the problem. A great value for $M$ is needed to associate a high energy to those configurations of the variables that do not respect the condition we are enforcing. In this way, we lower the probability for the quantum annealing process to produce configurations that violate the conditions.

\subsection{Layer 5 -- Algorithms: AQC as a sampler}

\subsubsection{Exploiting decoherence}

    Despite the fact that high level algorithms are not directly involved in some sort of compiling or control, we notice that there is a special use of the AQC which falls into the Layer 5, but it still has something to say about programming the stack to control some properties at the physical level, such as generating a statistics with some relevant properties to be employed in other algorithms. 
    Talking about applications of AQCs, we saw how such devices can be used to solve optimization problems. Unfortunately, they are not flawless in performing this task. Inside the QPU there are many sources of noise that interfere with the quantum annealing process (for a deep dive into the errors present in AQCS, see Ref. \cite{dwave2019technical}). Such errors cause decoherence and the resulting production of configurations different from the global minimum of the system.
    Researchers have tried to exploit such behaviour to use the AQC to produce configurations distributed according to a Boltzmann distribution \cite{denil2011toward}, \cite{dumoulin2014challenges}. Such phenomenon has been introduced in Sec. \ref{sec:Behaviour of multi-qubit chains}. It is expected that the sampling problem scales better on AQC with respect to classic processors, due to the ability of the quantum computer to simulate thermalization in the binary variables in a finite amount of time, independent of the dimension of the problem.
    
    Unfortunately, the effective temperature $T_\text{eff}$ in Eq. \ref{eq:boltzhypothesischapter4} changes quite rapidly and is difficult to forecast. Whenever the temperature changes, the sampling distribution produced by the AQC is affected. Whoever wants to use an AQC for sampling application has to know how to compensate such effect by rescaling the problem.

\subsubsection{Rescaling the problem} 
    \label{sec:rescaling weights}
    
    Suppose to define a generic problem Hamiltonian $H_P$, to be used during annealing, with weights and biases $J_{ij},\;h_i$. We then introduce:
    \begin{equation}
    \label{eq:rescaleweightsfordwave}
    \begin{split}
        &J^\prime_{ij} = J_{ij}\cdot \alpha \\
        &h^\prime_i = h_i\cdot\alpha
    \end{split}
    \end{equation}
    where $\alpha>0$ is a non negative parameter. It follows that
    \begin{equation}
    E(J^\prime_{ij},h^\prime_i, \{s_i\})=E(J_{ij}\cdot \alpha ,h_i\cdot \alpha , \{s_i\})=E(J_{ij},ah_i, \{s_i\})\cdot \alpha \;,
    \end{equation}
    where $E$ is the energy eigenvalue with respect to the state $\{s_i\}$ for the Hamiltonian $H_P$. The equation \ref{eq:rescaleweightsfordwave} holds because $H_P$ is homogeneous in weights and biases.
    
    Suppose that we implement the problem $H_P$ on the AQC and we obtain a resulting distribution with temperature $T'$. To produce correctly distributed samples it is, therefore, necessary to rescale weights and biases as in Eq. \ref{eq:rescaleweightsfordwave}, with $\alpha = T'$. Doing so, samples are extracted with a probability
    \begin{equation}
        P(\{s_i\})\propto e^{-E(J_{ij},h_i, \{s_i\})\frac{1}{T'}}=e^{-E(J_{ij},h_i,\{s_i\})\frac{\alpha }{T'}}=e^{-E(J_{ij},a_i,\{s_i\})}\;.
    \end{equation}
    
    If weights are rescaled before passing them to the QA processor, we can control the temperature of the final distribution, making it $\approx 1$. This technique is applied, e.g., in the field of quantum-trained Boltzmann Machines \cite{benedetti2016estimation}. Obviously, such approach can be applied only if the user knows how to estimate $T_\text{eff}$: a technique to do so is presented in Ref. \cite{benedetti2016estimation}.

\section{Commercial solutions for quantum firmware and quantum compiling}

Software developers around the world are starting to consider quantum computing as an innovative tool for solving many decades-old problems.
However, like classical computers and more specifically in analogy with FPGAs, it is not possible to ignore the characteristics of the hardware in use if one wants to obtain a result of any kind.
Therefore, quantum software developers require a wide range of skills and in particular a background in Physics is recommended. Commercial initiatives have started to fill the gap between quantum computers and developers without a background in quantum physics. This is particularly urgent in the current Noisy Intermediate Scale Quantum (NISQ) era \cite{preskill2018quantum}. Here, intermediate scale refers to a number of qubits ranging from 50 to 100 qubits.\\
The combination of the current number of qubits and the noise prevents the achievement of fault-tolerant quantum computers, a long term target planned for the next years.
For this reason, we may also expect that the tools needed for quantum developers will change as the NISQ era evolves to the fault-tolerant quantum computer.
Indeed, let's consider the quantum compiler as the tool to map a quantum algorithm on a register of logical qubits.
Since hundreds of physical qubits could be needed to have a logical qubit, and in the NISQ era only a few dozen are available, today a developer relies on a low level compiler to map an algorithm on a register of physical qubits. Such low level compilers cannot be hardware agnostic yet. Instead, developers needs an hardware-aware compiler.
For such reason, as anticipated, the problems of quantum compilation concerns various levels of abstraction \cite{haner2018software}, from high-level programming (Layer 4) to the pulse control of the qubits at the machine level (ayer 2).\\
In recent years, pulblic companies such as IBM, and startups like D-Wave and Rigetti have developed quantum computers and released open quantum programming languages, becoming the firsts quantum providers.
Such languages are software developer kits (SDKs) integrated in programming languages such as Python, C$++$, Java and C\# to allow professional developers to easily access the quantum resources made available by the provider.
Through such languages it is possible to map quantum algorithms on the physical qubits (not logical qubits to date) of a specific device yet many API and toolkits have been distributed for quantum software programming with a dual purpose:
\begin{itemize}
\settowidth{\leftmargin}{{\Large$\square$}}\advance\leftmargin\labelsep
\itemsep8pt\relax
\renewcommand\labelitemi{{\lower1.5pt\hbox{\Large$\square$}}}

\item Optimize the compilation of quantum algorithms 
\item Communicate with multiple providers.
\end{itemize}

Basically, the specific tools for quantum compilation available to users are hardware-aware compilers that allow you to compile your own algorithm on different technologies (superconductors, trapped ions, etc.) taking into account the typical noise of each device.

\subsection{Major players in the quantum compiler market}

We have already mentioned that several languages have been developed for quantum programming in the last years. In general many of such languages are Python-based quantum programming framework developed to allow the access to a specific hardware.
There are Qiskit and ProjectQ for the IBM backends, Cirq for the Google devices (which are not open for external users yet), pyQuil for Rigetti QPUs and Ocean for the D-Wave quantum adiabatic computers.\\
Recently Xanadu announced the release of the world's first available photonic quantum computer accessible with Strawberry Fields is a Python SDK that is based on the ‘continuous variable’ quantum computing paradigm.
There are also alternatives to Python-based quantum programming languages such as Q\# which is a hybrid classical–quantum language designed to facilitate the development of programs that can be run on a simulator.
Q\# is developed by Microsoft and it was designed to appeal to the C\# developer community.\\
Microsoft works in the development of a quantum computer with topological qubits and to date has no backends available for quantum computing users but through Azure cloud platform service it offers a series of quantum tools (Azure quantum) and an access to different quantum backends provided by other companies.
A similar strategy was undertaken by Amazon with AWS Braket providing access to a selection of backends from Rigetti, IonQ and D-Wave with its own Python-based language.
Through SDK Qiskit, IBM enables access to multiple technologies: its own computers with superconducting qubits and  the five-qubit trapped ion device at the University of Innsbruck (UIBK) hosted by Alpine Quantum Technologies (AQT).\\
Companies such as IBM, Rigetti and D-Wave have been able to develop a full-stack system from the backend to the user access providing a specific quantum compilers for their systems.
Over the years, specialized companies have partnered with full stack provider offering softwares (APIs and toolkits) to assist quantum developers by optimizing the compilation of their quantum algorithm.
Such specialized software companies like QCTRL and CQC (Cambridge Quantum Computing) offer tools at various levels of the compilation. \\
QCTRL, for example, is a IBM partner focused on the pulse control of the qubits. Recently IBM made available to users a framework for programming quantum circuits as a sequence of physical pulses for more accurate control over quantum transformations. In this context QCTRL offers toolkits to optimize the control and transformations on the backend qubits at the impulse level.
However, tools such as Boulder Opal \cite{ball2020software} are designed to be used by providers to make their devices programmable and is therefore a useful tool to enhance the firmware of quantum computers.\\
Cambridge Quantum Computing instead is a company specialized in quantum computing applications but also in the improvement of the compiling of the quantum algorithm at an intermediate level offering an hardware-aware compiler called t$\ket{ket}$ \cite{sivarajah2020t}.\\
In the next subsection we focus on the solutions made available in the main SDKs showing examples of the approach to quantum compilation adopted by the various providers to meet the needs of users.
In addition to this we will show t$\ket{ket}$ as an example of a tool made available by a third-party company to enhance the usability of quantum computers to traditional developers.

\subsection{Products for quantum compiling}
In the NISQ era various technologies and computational paradigms compete for primacy in the race to fully tolerant quantum computers. 
For such reason, a developer needs several compilers to use different qubit technologies.
Different native transformations, different sensitivity of the qubits to noise and different connection topologies between the qubits, a current compiler must take into account all of these hardware specific characteristics.
The most popular compilers are therefore specific to a given technology and there are still few cross-platform compilers. We see below the specifications of these compilers relating to the most popular programming languages \cite{larose2019overview}.

\textbf{Qiskit}. The Quantum Information Software Kit, or Qiskit, is
an open-source quantum software platform for working
with the quantum machine language, OpenQASM, of IBM quantum devices. Qiskit via host programming languages such as Python, JavaScript and Swift. 
The documentation of Qiskit can be found online at
\url{https://qiskit.org/documentation/}.\\ The documentation contains
instructions on installation and setup,
Qiskit overview, and developer documentation.
Qiskit is divided into 5 macro-modules: Terra, to build quantum circuits and compiler them enabling to run logical abstract circuit on an actual device, Aer, which provides optimized C++ simulator backends (also with a realistic noisy simulations) for executing circuits compiled in Terra, Ignis, used for the errors characterization of the devices and Aqua which includes a series of methods for real-world applications such as chemistry, optimization, finance and AI.\\
In terms of quantum computer architecture, Qiskit Terra is positioned as a compiler between the virtual and physical layers while Qiskit Aqua provides more abstract programming tools and therefore belongs to the application layer (quantum error correction and the logical layer are not yet implemented in NISQ devices).
Aer and Isign instead provide support tools for the benchmarking of software (through simulators) and devices (with calibration tools).
We will focus on the Qiskit Terra which provides the compiler for a quantum device.
In particular the a module of Qiskit Terra, called Trasnpiler allow the user to manage the compiler to set the accuracy level of the compilation.
Let's consider an example of a quantum circuit implemented with Qiskit

\begin{verbatim}
from qiskit import QuantumCircuit

qc = QuantumCircuit(5, 5)
qc.x(4)
qc.h(range(5))
qc.cx(range(0, 4), 4)
qc.h(range(5))
qc.measure(range(5), range(5))
\end{verbatim}
Two issues are to be taken into consideration for an optimal compilation on IBMQ systems:
\begin{itemize}
\settowidth{\leftmargin}{{\Large$\square$}}\advance\leftmargin\labelsep
\itemsep8pt\relax
\renewcommand\labelitemi{{\lower1.5pt\hbox{\Large$\square$}}}

\item The gate must be mapped into a basis of universal native gate of the device  
\item The qubits defined with the quantum circuit are \textit{virtual} representations of actual qubits used in computations. It is necessary to map these virtual qubits in a one-to-one manner to the physical qubits of the quantum device.
\end{itemize}
Qiskit Terra offers a fair level of customization of the compilation strategy.
The strategy to compile the quantum algorithms must be calibrated because a more efficient compilation requires a longer execution time but allows to obtain less noise-affected results.
In a nutshell the compiler executes three main operations: decompose all gates over three or plus qubits into only one or two qubits gates, map the gates into the native gates and map the virtual qubits into physical qubits.
Such operations are not necessarily in such order and other intermediate optimizations are performed by the qiskit compiler but the idea follows such main operations.
In the following, we show an example of how the transpiler module is used. Here, the backend taken into consideration is a simulator of an existing device called 'ibmq\_vigo'
\begin{verbatim}
from qiskit import transpile
from qiskit.test.mock import FakeVigo

backend = FakeVigo()
new_qc = transpile(qc, backend=backend, optimization_level=0)
\end{verbatim}
The transpiler module must know the specifics of the backend to use and the compilation efficiency can be set in optimization levels from $0$ to $3$ (in the example is $0$).
We will focus on the problem to map the virtual qubits (Layer 2) into the physical qubits.\\
The choice of mapping depends on the quantum circuit, the properties of the targeting device, and the optimization level that is chosen.
Such operation is performed by applying SWAP operations which map the input circuit onto the device topology and taking into account the noise properties of the device.
It should also be considered that each gate carries with it a considerable error, therefore a circuit consistent with the topology produces less noisy results since such a circuit requires a smaller number of SWAP gates.\\
However, we can customize the mapping strategy: We can set a trivial layout which maps virtual qubits to the same numbered physical qubit on the device or a dense layout where to obtain an optimal sub-graph of the device with same number of qubits of the circuit.
Optimization levels $0$ and $1$ have as default the trivial layout while $2$ and $3$ have the dense layout as default.\\
The choice of layout is of primary importance because superconductor computers have sparse connectivity and furthermore the entanglement between qubits on IBM computers is established through CNOT operations which are not bidirectional in general (in many pairs of qubits only one can be used as target and the other as control for a CNOT gate).


\textbf{pyQuil}. In 2017, Rigetti announced the first release of its quantum software platform developed called Forest which includes pyQuil, an open-source quantum programming language embedded in Python. pyQuil allows the users to build logical level quantum circuits in python and run them on Rigetti devices.
The compiler translate the logical quantum circuit into the quantum assembly language Quil \cite{smith2016practical}.
In the following example we have a simple random bit generator executed on the local quantum computer simulator called quantum virtual machine (QVM) with which it is possible to simulate 20-30 qubits on a normal CPU.
\begin{verbatim}
from pyquil.quil import Program
import pyquil.gates as gates
from pyquil import api

qp = Program()
qp += [gates.H(0) ,
        gates.MEASURE(0, 0)]
        
qvm = api.QVMConnection()
print(qvm.run(qprog), trials=1)
\end{verbatim}

\begin{figure}[]
\centering
\includegraphics[height=7.5cm]{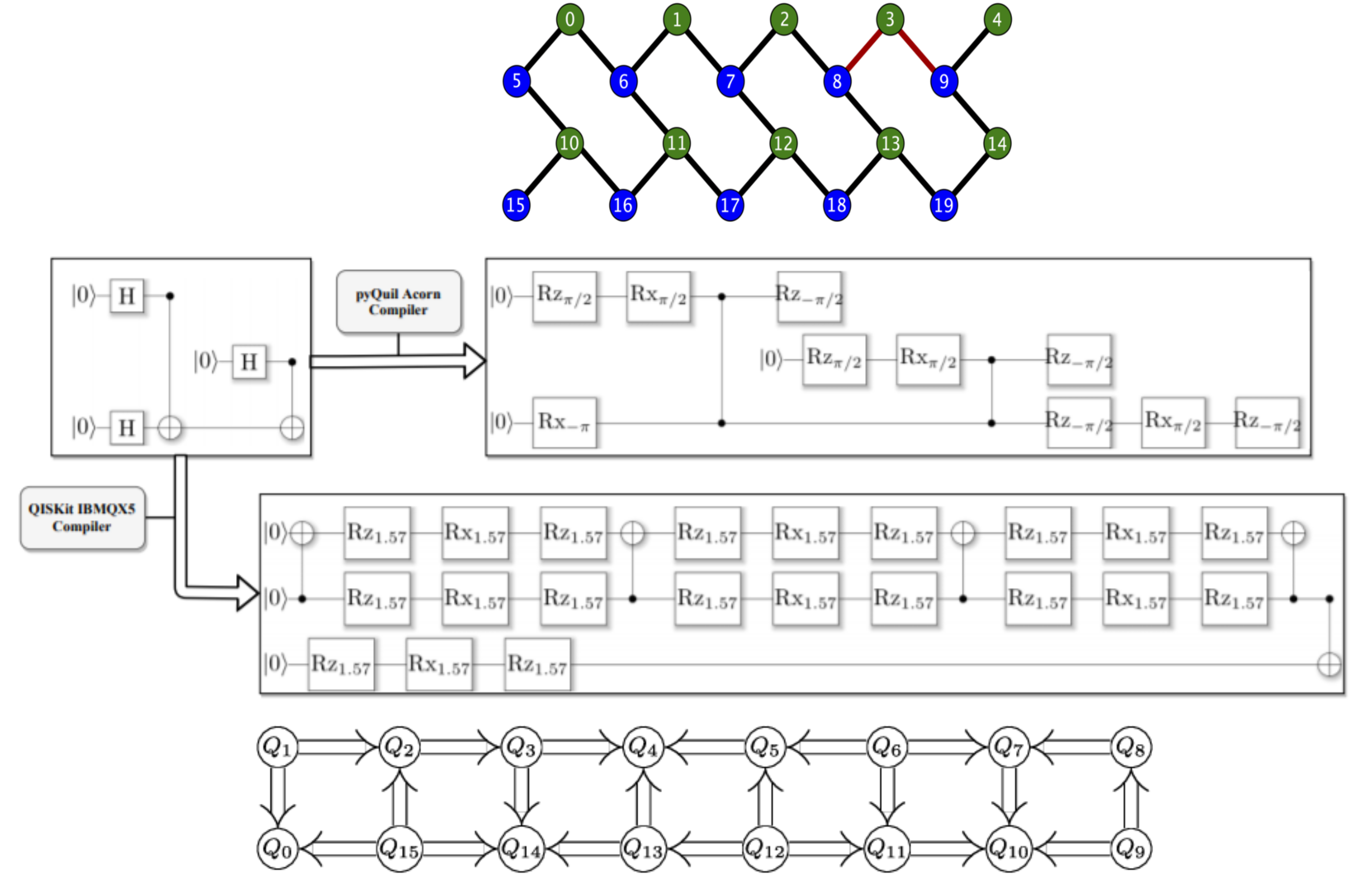}
\caption{Quantum circuit (Top left) compiled by pyQuil for Rigetti’s 8 qubit Agave processor and respective processor topology (top right), and by Qiskit for IBM’s 16 qubit IBMQX5 with the processor topology (down side).}
\label{fig:RigettiIBM}
\end{figure}

The instructions on installation and setup and the developer documentations can be found online at \url{https://docs.rigetti.com/}.
Rigetti devices, such as IBM ones, are quantum gate model computers with superconducting qubits and therefore the compilers, even if they use different strategies, they face the same problems as sparse connectivity between qubits. \\
The difference between the two devices is also in the native gates. It is important to note that Rigetti implements CZ gates on its devices instead of CNOT gates like IBM.
When we see the topology of a Rigetti device we must therefore consider that CZ is an invariant gate if two qubits are exchanged while this is not true for the CNOT which for each pair of qubits should be implemented in both directions.
We can see an example of the different action of a compilation of the same circuit into a Rigetti device and an IBM one in the Figure \ref{fig:RigettiIBM}.
Unlike the qiskit offer, in the case of pyquil there is not a high possibility to customize the compiler strategy.

\textbf{Ocean}. The suite of tools for D-Wave Systems is Ocean which provides a series of methods to map Ising and quadratic unconstrained binary optimization (QUBO) problems into the adiabatic quantum computers developed by D-Wave. 
Ocean is a python-based framework used to solve specific problem on QPUs and also on hybrid backends.
Ocean have a developer documentation with he instructions on installation and setup which can be found online at \url{https://docs.ocean.dwavesys.com/}.\\
The key concept is the embedding which is a map between the graph of the defined QUBO problem to the physical connection graph of the qubits in the quantum adiabatic devices. 
Today two connection graphs are been developed: Chimera and Pegasus. 
The first makes 6 connections per qubit while the second 15. Such connectivity is limited compared to the amount of qubits of these devices (thousands of qubits) due to the presence, also in this case, of noise.\\
The compiler made available on Ocean takes into account these limitations and provides a default level of embedding that can be improved through methods present in the framework. It is clear however that a more optimal embedding requires longer execution times to be found and this also depends the size of the problem under consideration.
Ocean provides a method called minorminer to set embedding accuracy level.
Given a minor and target graph, it tries to embed the minor into the target \cite{cai2014practical}.
Let's consider the following example
\begin{verbatim}
from minorminer import find_embedding

triangle = [(0, 1), (1, 2), (2, 0)]
square = [(0, 1), (1, 2), (2, 3), (3, 0)]

embedding = find_embedding(triangle, square, random_seed=10)
\end{verbatim}

\begin{figure}[]
\centering
\includegraphics[height=2.8cm]{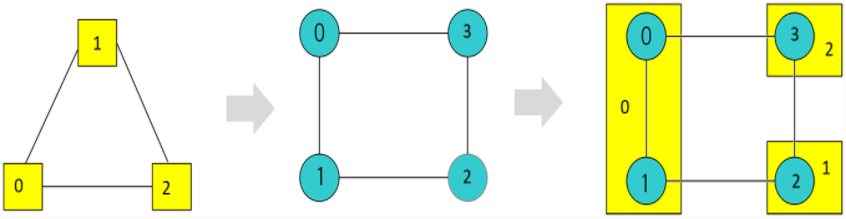}
\caption{Graphical representation of a minorminer operation which maps a triangle graph into a square target graph by chaining two target nodes to represent one source node.}
\label{fig:minorminer}
\end{figure}

In such example a triangle graph is mapped into a square graph. This embedding is necessary in many cases because, given the poor connectivity between qubits, it is sometimes necessary to add another qubit as a bridge. In the Figure \ref{fig:minorminer} it is graphically represented the problem.

\subsection{Future of the quantum compiling}
\begin{figure}[]
\centering
\includegraphics[height=7.2cm]{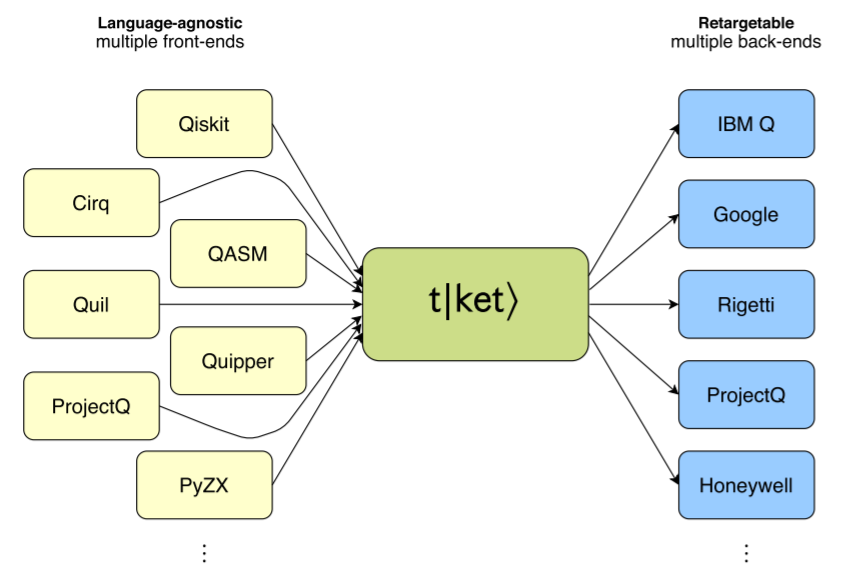}
\caption{Graphical representation of the concept of intermediate level quantum compiler. Algorithms written in different quantum programming languages such as Qiskit and Cirq, or quantum assembly languages such as QASM or Quil, can be executed into the backends of different providers using the pytket module which calls the C$++$ based t$\ket{ket}$ compiler. \cite{sivarajah2020t}}
\label{fig:tket}
\end{figure}

We have already discussed about how, in the NISQ era, the  quantum compilers cannot ignore the properties of the hardware since the challenge is precisely to increase the performance in terms of mitigating noise of these devices.
Concerning the future, we do not know whether the technologies such as ion trap or superconductor qubits will be able to scale to a device with hundreds of qubits, in order be merged in logical qubits, or to produce an adiabatic quantum computer with an high connectivity and low noise. Intel is working to achieve millions of silicon qubits at once in a wafer \cite{ferraro2020all}. 
For such reason truly agnostic compilers are investigated, focusing on an intermediate level of compilation.
One of the most promising examples is being developed by the quantum software company CQC (Cambridge Quantum Computing).
t$\ket{ket}$ is an open-source language-agnostic optimising compiler designed to run quantum algorithm on a variety of NISQ devices. 
It is composed by several features designed to minimise the
influence of device error.
While the core of t$\ket{ket}$ is a highly optimised C++ library, the system is available as the Python module
pytket.
The documentation is available online at \url{https://cqcl.github.io/pytket/}. 
Figure \ref{fig:tket} shows how the software can be interpreted as a intermediate level quantum compiler.
To interface with other software packages, and to use the relative backends, the user is supposed to install, in addition to pytket modules, also the plug-in packages:
pytket\_qiskit, pytket\_cirq, pytket\_pyquil, pytket\_projectq, and pytket\_pyzx.

\begin{verbatim}
from pytket import Circuit
from pytket.backends.ibm import IBMQBackend

circ = Circuit(4)
circ.CX(0, 1).CX(0, 2).CX(1, 2).CX(3, 2).CX(0, 3)

backend = IBMQBackend("ibmq_london")
backend.compile_circuit(circ)
handle = backend.process_circuit(circ, n_shots=2000)
\end{verbatim}
The example shows how pytket communicates with different backends 
By using the pytket, we can build a circuit which will be compiled to satisfy the constraints of the target backend, and then executed. To import the IBMQBackend class the pytket\_qiskit extension package is required. We can also import a circuit written in a different language and append operations using pytket
\begin{verbatim}
circuit = circuit_from_qasm('input.qasm')
q0, q1 = circuit.qubits[:2]
circuit = cirucit.H(q0).CX(q0, q1).measure_all()
\end{verbatim}
In such example a circuit is read in from a QASM file; and pytket is used to append other operations.
t$\ket{ket}$ is  one of the examples of a market, that of quantum computing, which is slowly coming out of the labs to spread among the developer communities.

\bibliography{bibliography}

\end{document}